\documentstyle[12pt,epsfig]{article}

\begin{document}

\begin{center}
{\Large Uncertainties and Discovery Potential in Planned Experiments}
\end{center}

\bigskip

\begin{center}
S.I.~Bityukov$^2$,
N.V.~Krasnikov$^1$\\

\bigskip

Institute for High Energy Physics,\\
142284, Protvino Moscow Region, Russia
\end{center}

\bigskip

\begin{abstract}
We describe a method for estimation of the discovery potential 
on new physics in planned experiments. 
The effective significance of signal for given probability of
observation is proposed for planned experiments instead of the usual 
significances $\displaystyle S_1 = \frac{n_s}{\sqrt{n_b}}$ and
$\displaystyle S_2 = \frac{n_s}{\sqrt{n_s+n_b}}$, where $n_s$ and 
$n_b$ are the average numbers of signal and background events.
Application of the test of equal-probability allows to estimate
the exclusion limits on new physics.
We also estimate the influence of systematic uncertainty related to nonexact 
knowledge of signal and background cross sections on the 
discovery probability of new physics in planned experiments.
An account of such systematics is very essential in the search for 
supersymmetry at LHC. 
\end{abstract}

\noindent
{\tt Keywords:  Poisson Distribition, uncertainties, hypotheses testing,
significance.} 

\bigskip

\vspace*{4cm}
\noindent
\rule{3cm}{0.5pt}\\
$^1$Institute for Nuclear Research RAS, Moscow, Russia\\
$^2$Email addresses: bityukov@mx.ihep.su, Serguei.Bitioukov@cern.ch

\newpage

\section{Introduction}

One of the common goals in the forthcoming experiments is the search for new 
phenomena. In the forthcoming high energy physics experiments 
(LHC, TEV22, NLC, ...) the main goal is the search for physics
beyond the Standard Model (supersymmetry, $Z'$-, $W'$-bosons, ...) and
the Higgs boson discovery as a final confirmation of the Standard Model.
In estimation of the discovery potential of the planned experiments
(to be specific in this paper we shall use as an example CMS experiment
at LHC~\cite{CMS}) the background cross section (the Standard Model
 cross section)
is calculated and for the given integrated luminosity $L$ the average 
number of background events is $n_b = \sigma_b \cdot L$.
Suppose the existence of a new physics 
leads to additional nonzero signal cross section $\sigma_s$ with the same 
signature as for the background cross section  that 
results in  the prediction
of the additional average number of signal events \footnote{It should be 
noted that the existence of new physics can also lead to the decrease 
of the cross section due to destructive interference or some nonlocal 
formfactors. In this paper we consider the case when the new physics 
existence leads to additional positive contribution to 
the background cross section. The consideration of the opposite case 
is straightforward.}   
$n_s = \sigma_s \cdot L$ for the integrated luminosity $L$.

The total average number of the events is 
$<n> = n_s + n_b = (\sigma_s + \sigma_b) \cdot L$.
So, as a result of new physics existence, we expect an excess 
of the average number of events. In real experiments the probability
of the realization of $n$ events is described by Poisson 
distribution~\cite{Frodesen,PDG}

\begin{equation}
f(n; \lambda)  = \frac{{\lambda}^n}{n!} e^{-\lambda}.
\end{equation}

\noindent
Here $\lambda = <n>$ is the average number of events. 
Remember that the Poisson distribution $f(n;\lambda)$ gives~\cite{Frodesen} 
the probability of finding exactly $n$ events in the given interval
of (e.g. space and time) when the events occur independently
of one another at an average rate  of $\lambda$ per the 
given interval. For the Poisson distribution the variance $\sigma^2$ equals 
to $\lambda$. So, to estimate the probability of the new physics discovery 
we have to compare the Poisson statistics with $\lambda = n_b$ and
$\lambda = n_b + n_s$. Usually, high energy physicists use the following  
``significances'' for testing the possibility to discover new physics
in an  experiment:

\begin{itemize}
\item[(a)] ``significance'' 
$S_1 = \displaystyle \frac{n_s}{\sqrt{n_b}}$~\cite{CMS,ATLAS},
\item[(b)] ``significance'' 
$S_2 = \displaystyle \frac{n_s}{\sqrt{n_s + n_b}}$~\cite{S2},
\item[(c)] ``significance'' 
$2 \cdot S_{12} = 
2 (\displaystyle \sqrt{n_s + n_b} - \sqrt{n_b})$~\cite{Brown,Bit1}.
\end{itemize}

\noindent
A conventional claim is that for $S_1~(S_2) \ge 5$ we shall discover
new physics (here, of course, the systematic uncertainties are ignored).
For $n_b \gg n_s$ the significances  $S_1$ and $S_2$ coincide (the 
search for Higgs boson through the $h \rightarrow \gamma \gamma$
signature). For the case when $n_s \sim n_b$, ~$S_1$ and $S_2$ differ.
Therefore, a natural question arises: what is the correct definition for the 
significance $S_1$, $S_2$ or anything else ?

It should be noted  that there is a crucial difference between the planned
experiment and the real experiment. In the real experiment the total 
number of events $n_{obs}$ is a given number (already has been measured)
and we compare it with $n_b$ when we test the validity of the
standard physics. So, the number of possible signal events is determined
as $n_s = n_{obs} - n_b$ and it is compared  with the average number of 
background events $n_b$. The fluctuation of the background is 
$\sigma_{fb} = \sqrt{n_b}$, therefore, we come to the $S_1$ significance
as the measure of the distinction from the standard physics.
In the conditions of the planned  experiment when we want to search
for new physics, we know only the average number of the background
events and the average number of the signal events, so we
have to compare the Poisson distributions $f(n; n_b)$ and
$f(n; n_s + n_b)$ to determine the probability to find new physics
in planned experiment.

In this paper we describe a method for estimation of the discovery potential 
and exclusion limits on new physics in planned experiments. 
The effective significance of signal for given probability of
observation is proposed for planned experiments instead of the usual 
significances $\displaystyle S_1 = \frac{n_s}{\sqrt{n_b}}$ and
$\displaystyle S_2 = \frac{n_s}{\sqrt{n_s+n_b}}$, where $n_s$ and 
$n_b$ are the average numbers of signal and background events.
We also estimate the influence of systematic uncertainties 
related to nonexact knowledge of signal and background cross sections 
on the probability to discover new physics in planned experiments. 
An account of such systematics is very essential in the search for 
supersymmetry at LHC.

The organization of the paper is the following. 
In the next Section mainly due to completeness we 
discuss  the case of real experiment.
In Section 3 we describe a method for 
the estimation of   new physics discovery potential 
in planned experiment. Section 4 deals with estimation of exclusion limit. 
In Section 5  we estimate the influence of the systematics related to nonexact 
knowledge of the signal and background cross sections on the probability 
to discover new physics and set up exclusion  limits on new physics in 
planned experiments. Section 6 contains concluding remarks.

\section{New physics discovery in real experiment}

In this section well known 
situation with real experiment is reminded to pedagogical reasons. 
Consider the case when the average number  $\lambda$ of the events 
in the Poisson distribution (1) is big $\lambda \gg 1$. 
In this case the Poisson distribution (1) approaches the 
Gaussian distribution 
  
\begin{equation}
f_G(n; \mu,\sigma) = \int_{n-0.5}^{n+0.5}{P_G(x; \mu,\sigma^2)dx}, 
\end{equation}

\noindent
with $P_G(x; \mu,\sigma^2) = \frac{1}{\sigma \sqrt{2 \pi}} \cdot
e^{-\frac{(x - \mu)^2}{2 \sigma^2}}$, $\mu = \sigma^2$, $\mu = \lambda$ 
and $n \ge 0$. 
Note that for the Poisson distribution the mean equals to the variance. 
According to common definition \cite{PDG} new physics discovery 
corresponds 
to the case when the probability that background can imitate signal is 
less than $5 \sigma$ (here of course we neglect any possible systematic 
uncertainties). Suppose we have observed some excess of events $n_{obs} >n_b$.
The probability that for the background $n_b$ we shall observe 
events with $n \geq n_{obs}$ is determined by standard formula
\begin{equation}
P(n \geq n_{obs}) = \int^{\infty}_{n_{obs}}P_G(x; n_b,n_b)dx =
\frac{1}{\sqrt{2\pi}}\int^{\infty}_{S_1}exp(-x^2/2)dx ,
\end{equation}
where 
\begin{equation}
S_1 =\frac{n_{obs}-n_b}{\sqrt{n_b}} \equiv \frac{n_s}{\sqrt{n_b}}
\end{equation}
According to common definition for $S_1 \geq 5$ the Standard Model 
is excluded (the probability that background can imitate signal is less than
$2.85 \cdot 10^{-7}$) and we have new physics discovery \footnote{Here we 
neglect any possible systematic uncertainties.}. 

Suppose 
some model with new physics predicts $\lambda = n_s + n_b$ and
\begin{equation}
|n_{obs} - n_s - n_b| \leq k_1 \cdot  \sigma_{s + b},
\end{equation}
then for $k_1 = 1.64$ ($k_1 = 1.96$) the model with ``new physics'' 
agrees with experimental data at 90\% C.L. ( 95\% C.L.). 
Here $\sigma^2_{s+b} = n_s + n_b$.
  
For $S_1 \geq 2$ ($S_1 \geq 3$ ) in formula (3)  
the probability that background can 
imitate signal is less than 2.28\% ( 0.14\%) and according to 
\underline{our definition} 
we have weak (strong) evidence in favor of new physics. 

Suppose that the measured number of events is such that 
\begin{equation}
| n_{obs} - n_b| \leq (k_1 = 1.96) \cdot (\sigma_{b} = \sqrt{n_b})
\end{equation}
It means that the Standard Model agrees at 95 \% C.L. 
with experimental 
measurement. In this case we can also obtain  exclusion limit 
 on new physics (limit on the average number of signal events $n_s$) . 
Namely, for $\lambda~=~n_s~+~n_b$ we require that 
\begin{equation}
| n_{obs}-n_s-n_b | \leq (k_1=1.96) \cdot (\sigma_{s+b} = {\sqrt{n_s + n_b}})  
\end{equation}
From the equation (7) we obtain 95\% C.L. upper bound on the 
average number  $n_s$ of signal events.
    
Consider now the case of the Poisson distribution (1). Suppose we have 
measured the number of events $n_{obs} > n_b$ (an excess of events). 
We define the statistical significance $s$ of a signal~\cite{Narsky} 
in the Standard Model by

\begin{equation}
\frac{1}{\sqrt{2\pi}}\int_{s}^{\infty}exp(-x^2/2)dx =
\sum_{k = n_{obs}}^{\infty}f(k; n_b).
\end{equation}   

\noindent
The formula (8) is nothing but the probability 
to observe  $n \ge n_{obs}$ of background events in an identical
independent experiments. Note that $s$ is a 
function on $n_{obs}$ and $n_b$,~ $s = s(n_{obs},n_b)$. 
If $s \geq 5$ then  by common definition we have 
new physics discovery. For $s \geq 2$ ($s \geq 3$) 
according to our definition we have weak(strong) evidence in favor of 
new physics. If the model with additional  
$n_s$ signal events obeys the inequality

\begin{equation} 
 |s(n_{obs}, n_s +n_b)| \leq (k_1 = 1.96)   
\end{equation}

\noindent
then the  model with new physics agrees at 95\% C.L. with an experiment.

Suppose that we have observed the number of events compatible at 
95\% C.L. with the Standard Model, i.e.

\begin{equation}
s(n_{obs}, n_b) \leq (k_1 = 1.96).            
\end{equation}

\noindent
In this case one can obtain at 95\% C.L. 
exclusion limit  on the average number of 
signal events $n_s$ from the inequality

\begin{equation}
s(n_{obs}, n_s +n_b) \leq (k_1 = 1.96).
\end{equation}
 
\section{Planned experiments.}

As it has been mentioned in the introduction the crucial difference 
between planned experiment and real experiment is that in real experiment we 
know the number of observed events, 
therefore we can compare the Standard Model 
with experimental data directly, whereas in the case of planned experiment 
we know only the average number of background events $n_b$ and the 
average number of signal events (for the case when we have new physics 
in addition to the Standard Model). Therefore in the case of planned 
experiment an additional ``input'' parameter is the probability of 
the discovery. Suppose we test two models: the Standard Model with 
the average number of events $\lambda = n_b$ and the model with new physics 
and  the average number of events $\lambda = n_s +n_b$. 

To discover new physics we have to  require that the 
probability $\beta(\Delta)$ of the background fluctuations for 
$n > n_0(\Delta)$ is less than $\Delta$, namely 
\begin{equation}
\beta(\Delta) =  \sum ^{\infty}_{n=n_0({\Delta})+1} f(n; n_b) \leq \Delta
\end{equation}

\noindent
The probability $1 - \alpha(\Delta)$
that the number of  events in a model with new physics will be bigger than 
$n_0(\Delta)$ is equal to 
\begin{equation}
1 - \alpha(\Delta) = \sum ^{\infty}_{n = n_0(\Delta) + 1}
f(n; n_s + n_b)
\end{equation}

\noindent
It should be stressed that if $\Delta$ is a given 
number then $\alpha(\Delta)$ 
is a function of $\Delta$ or vice versa we can fix the value of 
$\alpha$ in formula (13) then $\Delta$ is a function of $\alpha$.
The meaning of the probability of the discovery $1 - \alpha$ is 
the probability that 
in the case of new physics an experiment will measure the number 
of events bigger than $n_0$ such that the probability that the Standard 
Model can reproduce such number of events is rather small ($\beta$). 

In other words we choose the critical value $n_0$ for hypotheses
testing\footnote{A simple statistical hypothesis $H_0$ 
(new physics is present, 
i.e. $\lambda = n_s + n_b$) against a simple alternative hypothesis
$H_1$ (new physics is absent, i.e. $\lambda = n_b$)~\cite{Frodesen}.}
about observability of new physics requiring that 
Type II error $\beta \le \Delta$. Then we calculate the 
Type I error $\alpha$ and the probability of discovery
(or evidence) $1 - \alpha$.

For fixed 
value of $\alpha$ and known values of $n_s$, $n_b$ we can calculate 
$\beta$ using formulae (12,13). In our numerical calculations we take 
$\alpha = 0.5;~0.25;~0.1;~0.05$. Consider now the 
limiting case $n_b \gg 1$ when Poisson distribution approaches Gaussian 
distribution. The  equations (12,13) take the form 

\begin{equation}
\beta \approx \int_{n_0}^{\infty}{P_G(x; n_b, n_b)dx}
\end{equation}

\begin{equation}
1 -\alpha  \approx \int_{n_0}^{\infty}{P_G(x; n_s +n_b, n_s + n_b)dx}
\end{equation}
 
Consider at first  the most simple case 
when $\alpha = 0.5$ (see Figs.1-2 for an 
illustration). For $\alpha = 0.5$  parameter $n_0$ in formula (15) 
is equal to 
$n_0 = n_s + n_b$. The equation (14) takes the form    
 
\begin{equation}
\beta \approx \int_{S_1}^{\infty}{P_G(x; 0, 1)dx},
\end{equation}
where 
\begin{equation}
S_1 = \frac{n_s}{\sqrt{n_b}}
\end{equation}

\begin{figure}[htpb]

  \begin{center}
 \epsfig{file=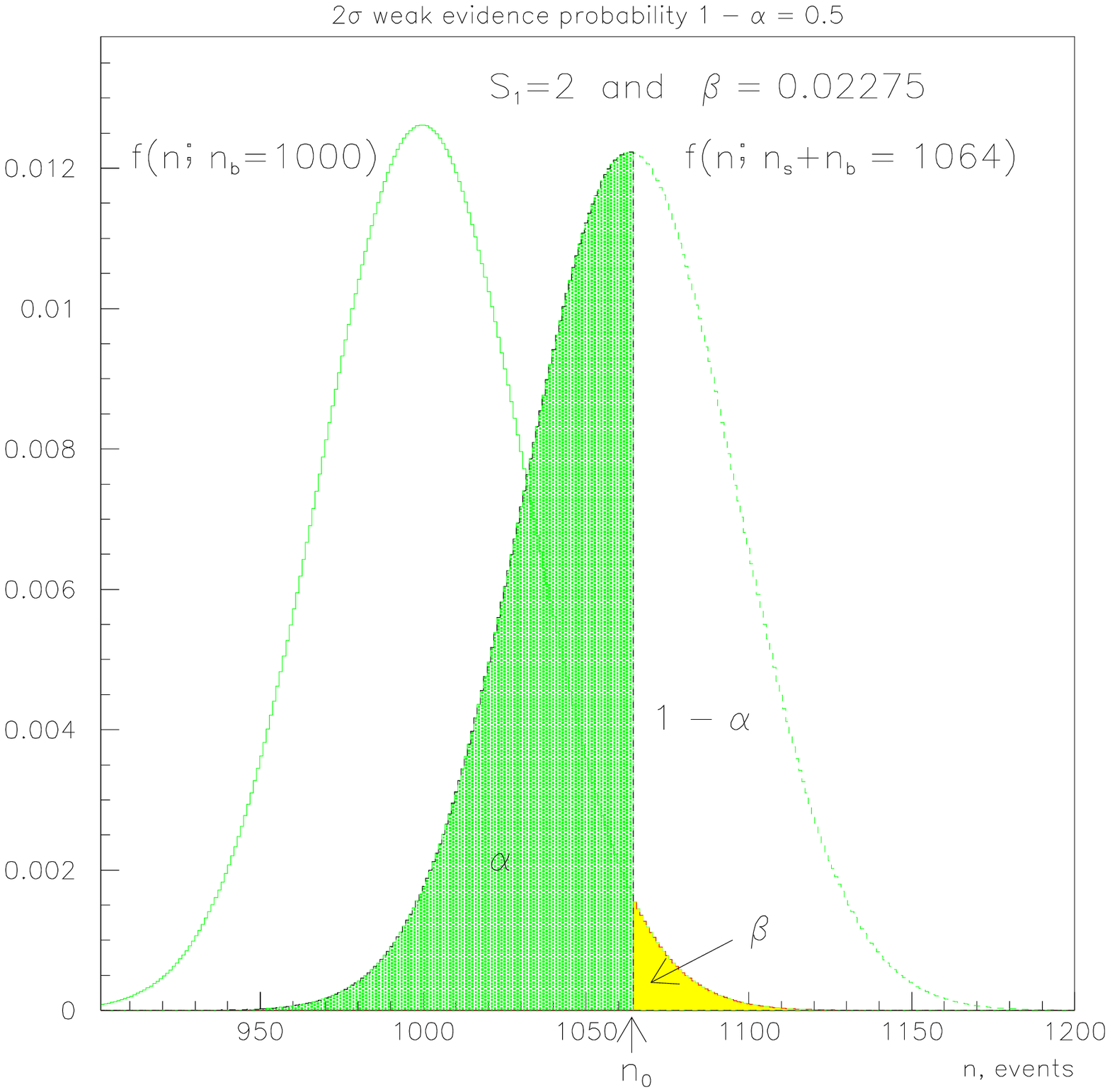,width=6.2cm}
\caption{The case $n_b \gg 1$. Poisson distributions with
parameters $\lambda = 1000$ (left) and $\lambda = 1064$ (right). Here 
$1 - \alpha = 0.5$ and $\beta = 0.02275$ (i.e. $S_1 = 2$).}
    \label{fig:1} 
  \end{center}
                
  \begin{center}
 \epsfig{file=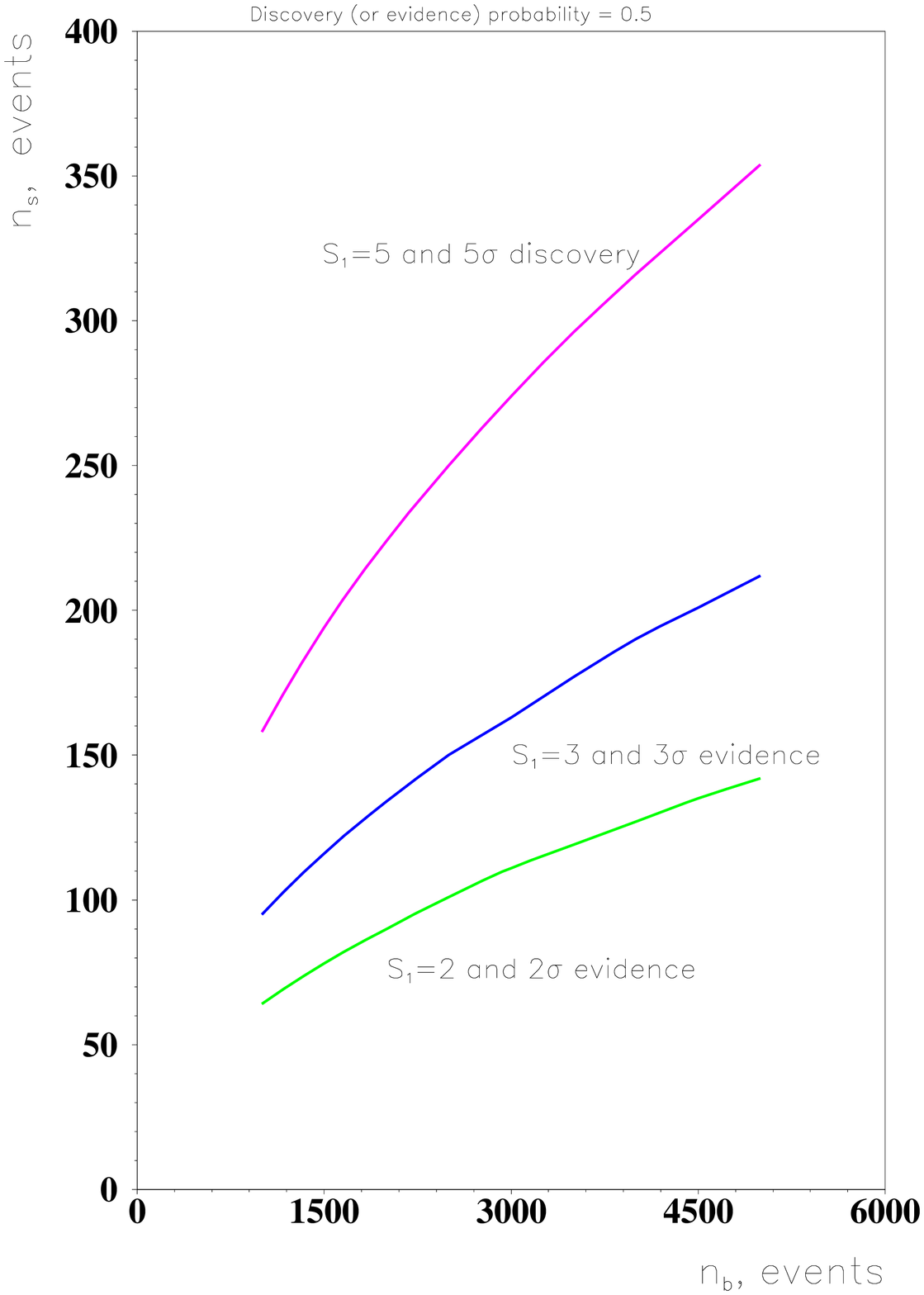,width=6.2cm}
\caption{The case $n_b \gg 1$. Dependences $n_s$ versus $n_b$ for 
$S_1 = 5$, $S_1 = 3$ and $S_1 = 2$ coincide with $5\sigma$ discovery, 
$3\sigma$ strong evidence, and $2\sigma$ weak evidence curves, 
correspondingly. The probability of discovery $1 - \alpha = 0.5$.}
    \label{fig:2} 
  \end{center}
\end{figure}

\noindent
The significance $S_1$ is determined by the formula (17) and it is
often used in experiment proposals~\cite{CMS,ATLAS}. 

For  $1 - \alpha > 0.5$ (see Fig.3 for an illustration) 
the  parameter $n_0$ in formula (14) is equal to 
\begin{equation}
n_0 = n_s + n_b - k(\alpha)\sqrt{n_s + n_b}, 
\end{equation}
where $k(\alpha)$: 
$k(0.5) = 0$; $k(0.25) = 0.66$; $k(0.1) = 1.28$; $k(0.05) = 1.64$ 
(as an example, Tab.28.1~\cite{PDG}). 
The effective significance $s$ in the equation (8) 
(i.e. corrected significance $S_1$, corresponding the discovery 
probability $1-\alpha$) has the form
\begin{equation}
s =\frac{n_s}{\sqrt{n_b}} - k({\alpha})\sqrt{1 + \frac{n_s}{n_b}}
\end{equation}

\begin{figure}[htpb]
  \begin{center}
 \epsfig{file=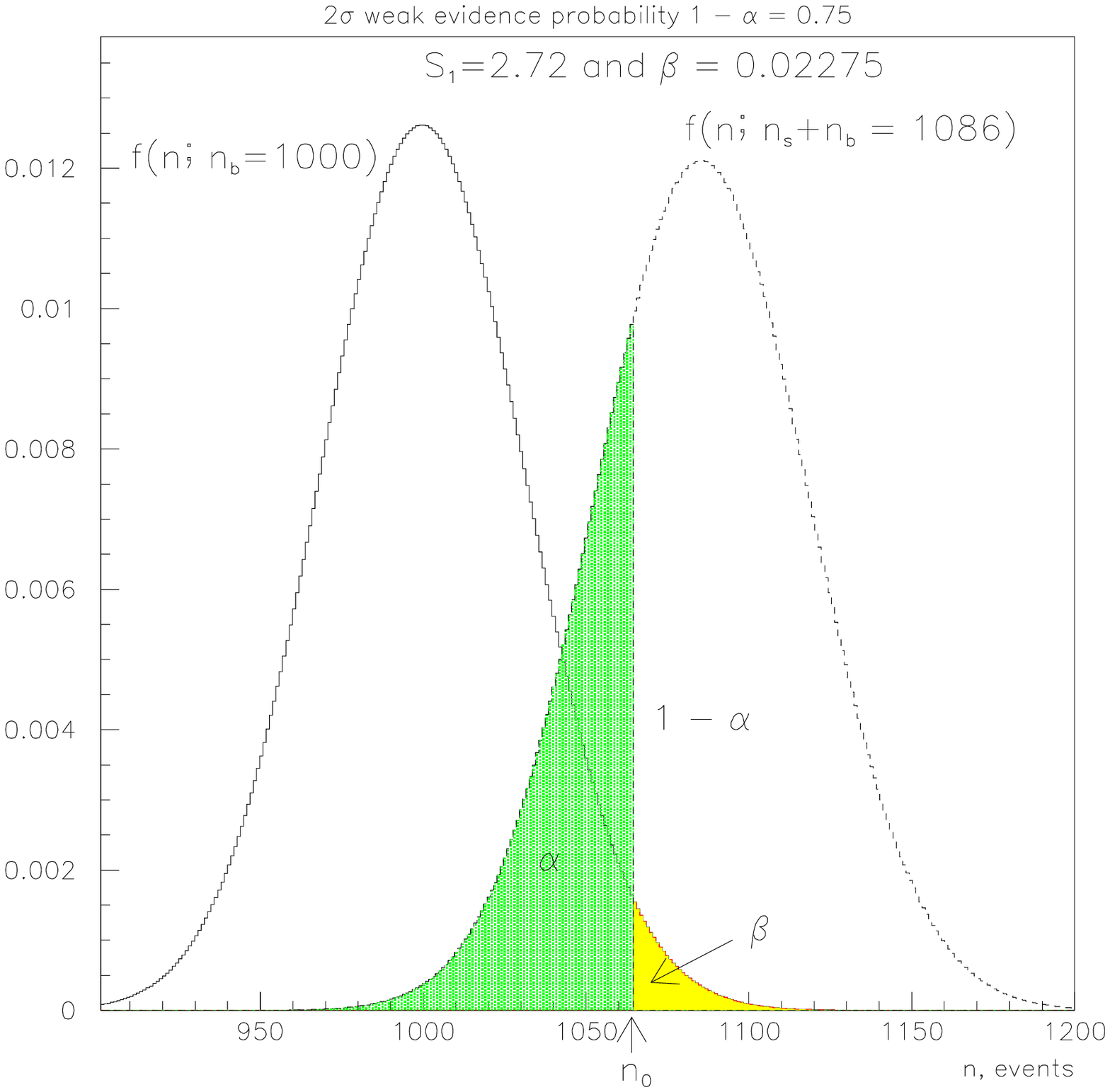,width=6.2cm}
\caption{The case $n_b \gg 1$ and $S_1 = 2.72$. Poisson distributions with
parameters $\lambda = 1000$ (left) and $\lambda = 1086$ (right). Here 
$1 - \alpha = 0.75$ and $\beta = 0.02275$ (i.e. effective $s = 2$).}
    \label{fig:3} 
  \end{center}

  \begin{center}
   \epsfig{file=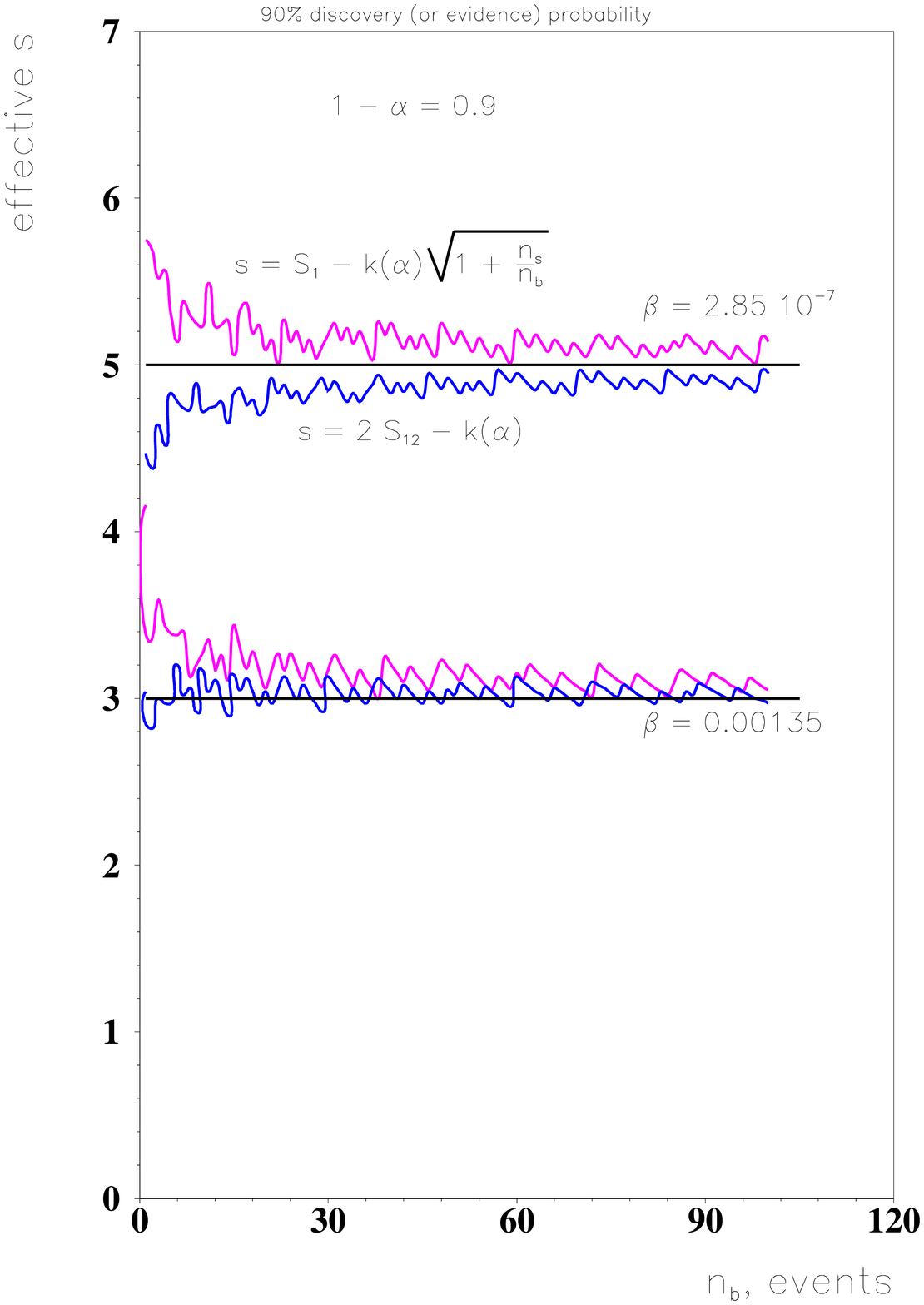,width=6.2cm}
\caption{The estimation of effective significance $s$ 
for given $\beta$ and $1 - \alpha$.}
    \label{fig:4} 
  \end{center}
\end{figure}
                 
\noindent                
So, we see that  the asymptotic formula (17) for the significance $s$ is 
valid only for $1 - \alpha = 0.5$. 
 
As it has been shown in refs.~\cite{Bit2,Bit4}
the more proper of the significance in planned experiments is 
$2 \cdot S_{12}$. The generalization of this significance to the case of 
$1 - \alpha > 0.5$ looks very attractive for approximate estimation of
discovery potential
\begin{equation}
s = 2 \cdot (\sqrt{n_s+n_b} - \sqrt{n_b}) - k({\alpha}).
\end{equation}

\noindent                
The comparison of formulae (19,20) is shown in Fig.4.

It should be stressed that very often in the conditions of planned 
experiment the average numbers of background and real events are 
not very big and we have to solve the equations (12, 13)
directly to construct $5\sigma$ discovery, $3\sigma$ strong evidence
and $2\sigma$ weak evidence curves. 
Our numerical results are presented in Figs. (5 - 10).  

\begin{figure}[htpb]
  \begin{center}
 \epsfig{file=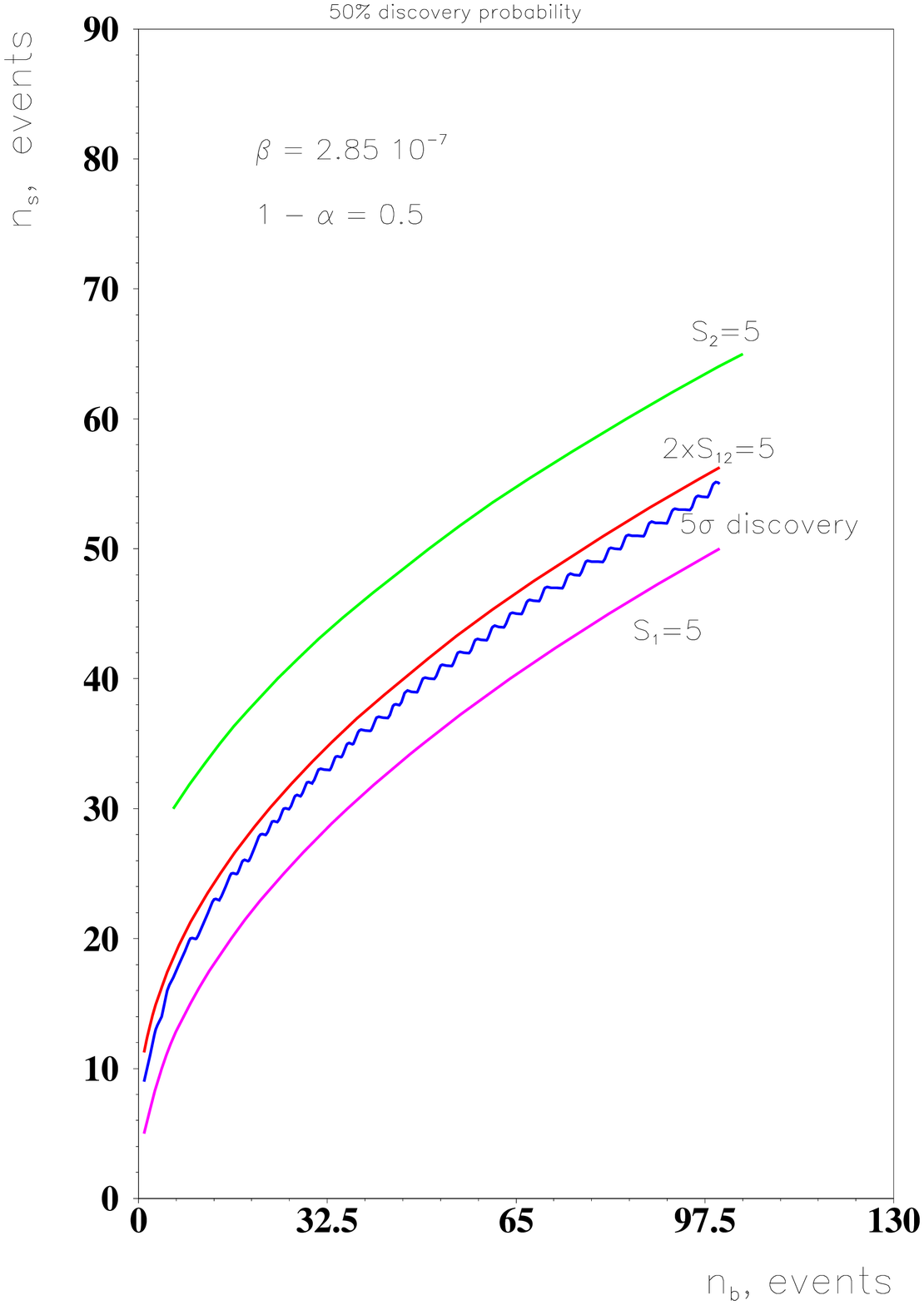,width=12.0cm}
\caption{The $5 \sigma$ discovery curve and dependences $n_s$ versus 
$n_b$ for $S_1 =5$, $S_2 = 5$, $~2 \cdot S_{12} = 5$.
Here $1 - \alpha = 0.5$ and $\beta = 2.85 \cdot 10^{-7}$.}
    \label{fig:5} 
  \end{center}
\end{figure}
                
\begin{figure}[htpb]
  \begin{center}
  \epsfig{file=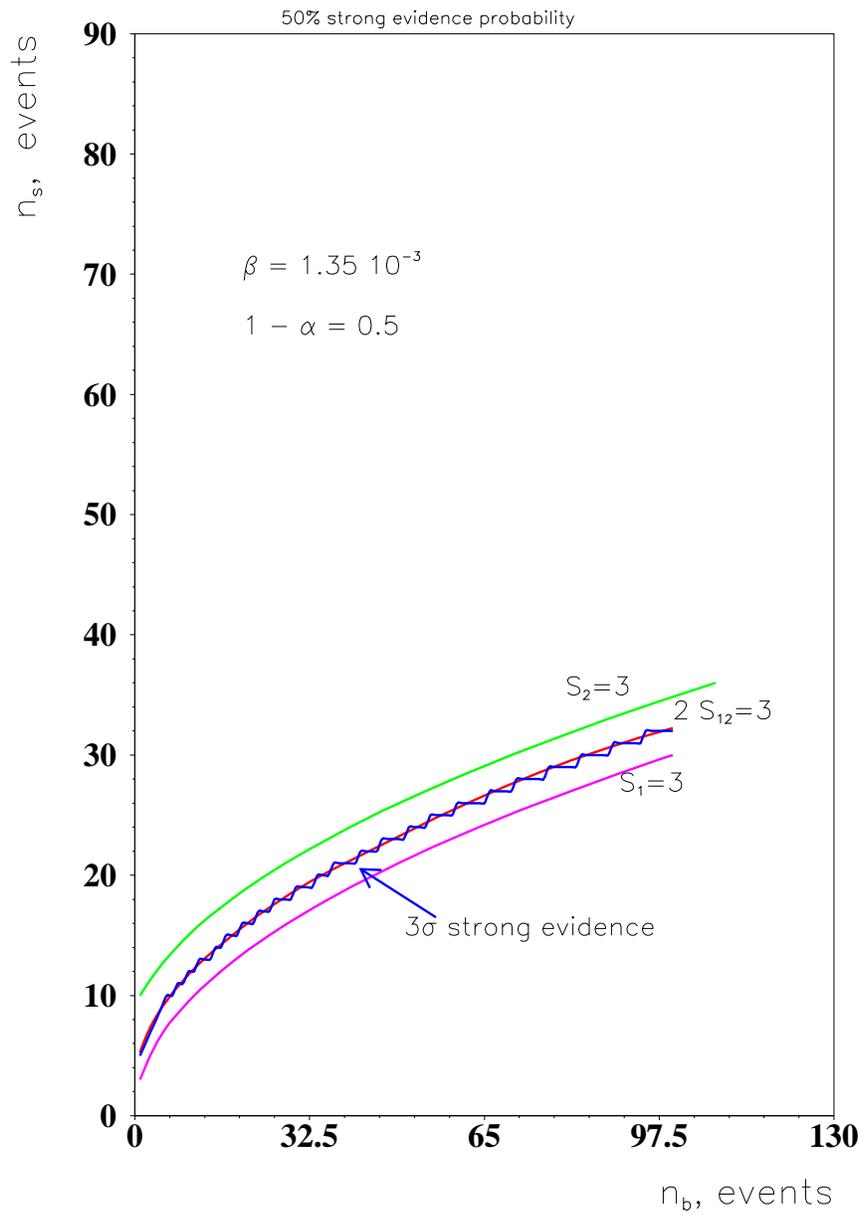,width=12.0cm}
\caption{The $3 \sigma$ strong evidence curve and dependences $n_s$ versus 
$n_b$ for $S_1 =3$, $S_2 = 3$, $~2 \cdot S_{12} = 3$.
Here $1 - \alpha = 0.5$ and $\beta = 0.00135$.}
    \label{fig:6} 
  \end{center}
\end{figure}
                
\begin{figure}[htpb]
  \begin{center}
  \epsfig{file=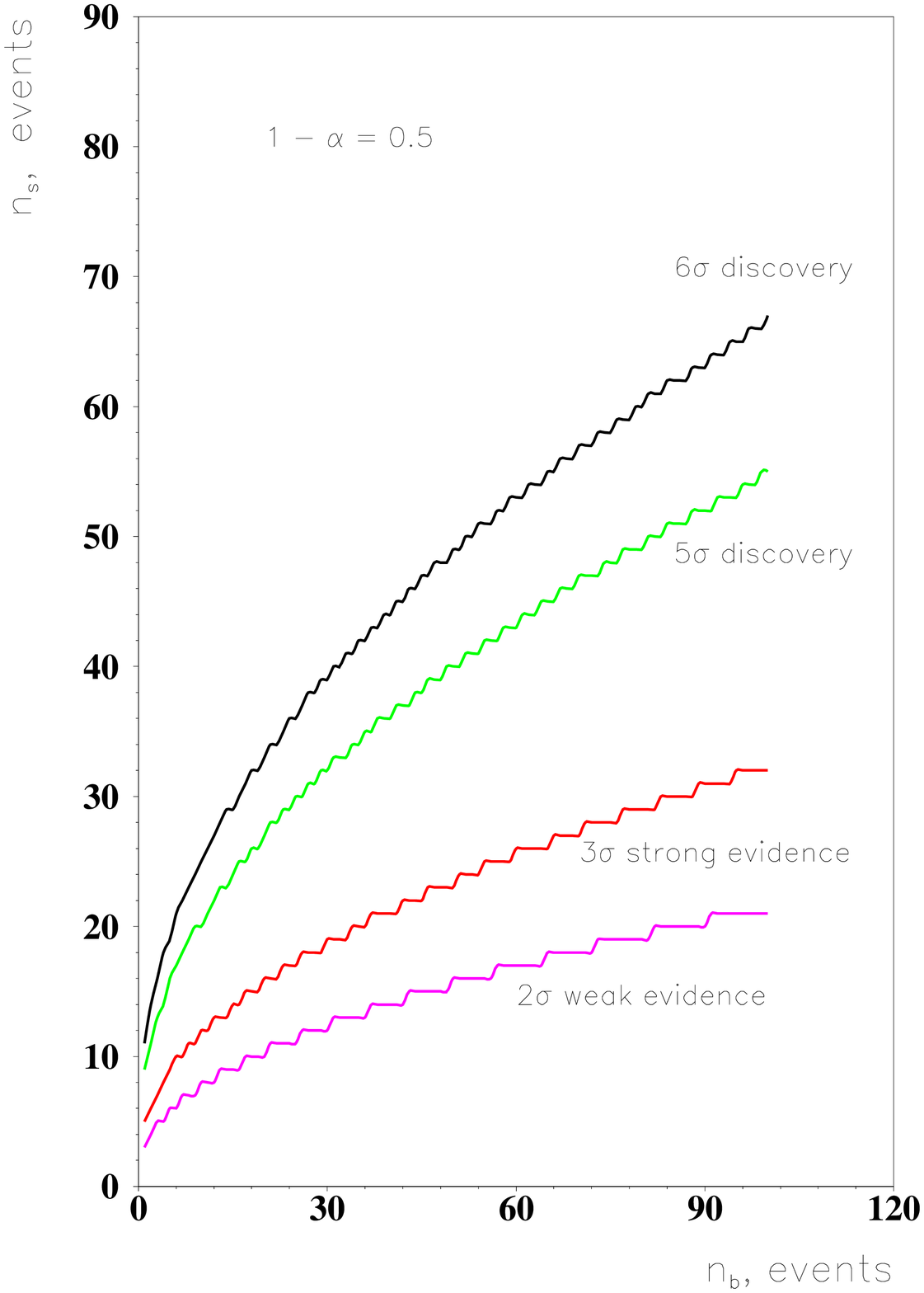,width=12.0cm}
\caption{Dependences $n_s$ versus $n_b$ for
$1 - \alpha = 0.5$ and for different values of $\beta$.}
    \label{fig:7} 
  \end{center}
\end{figure}
                
\begin{figure}[htpb]
  \begin{center}
  \epsfig{file=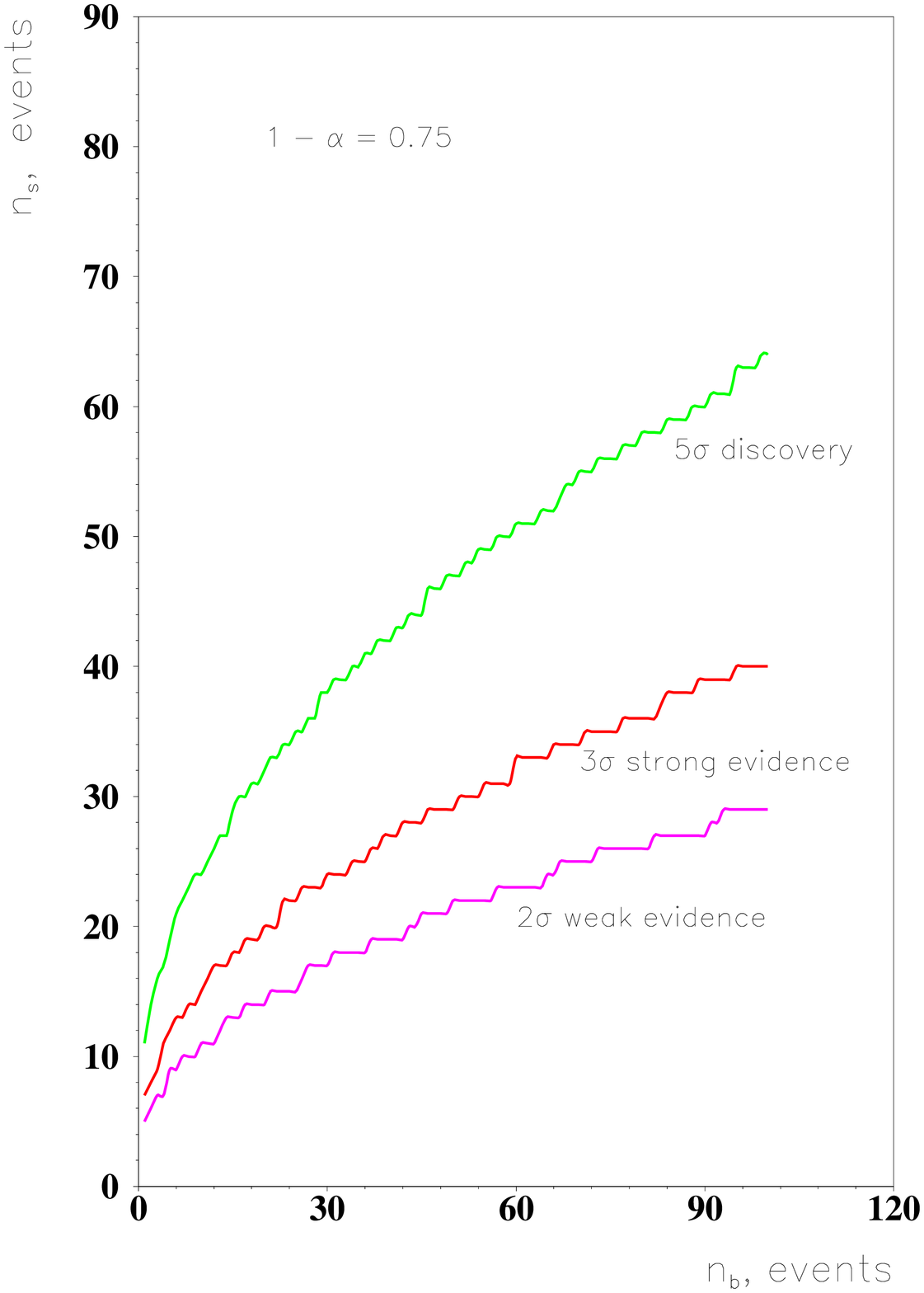,width=12.0cm}
\caption{Dependences $n_s$ versus $n_b$ for
$1 - \alpha = 0.75$ and for different values of $\beta$.}
    \label{fig:8} 
  \end{center}
\end{figure}
                
\begin{figure}[htpb]
  \begin{center}
  \epsfig{file=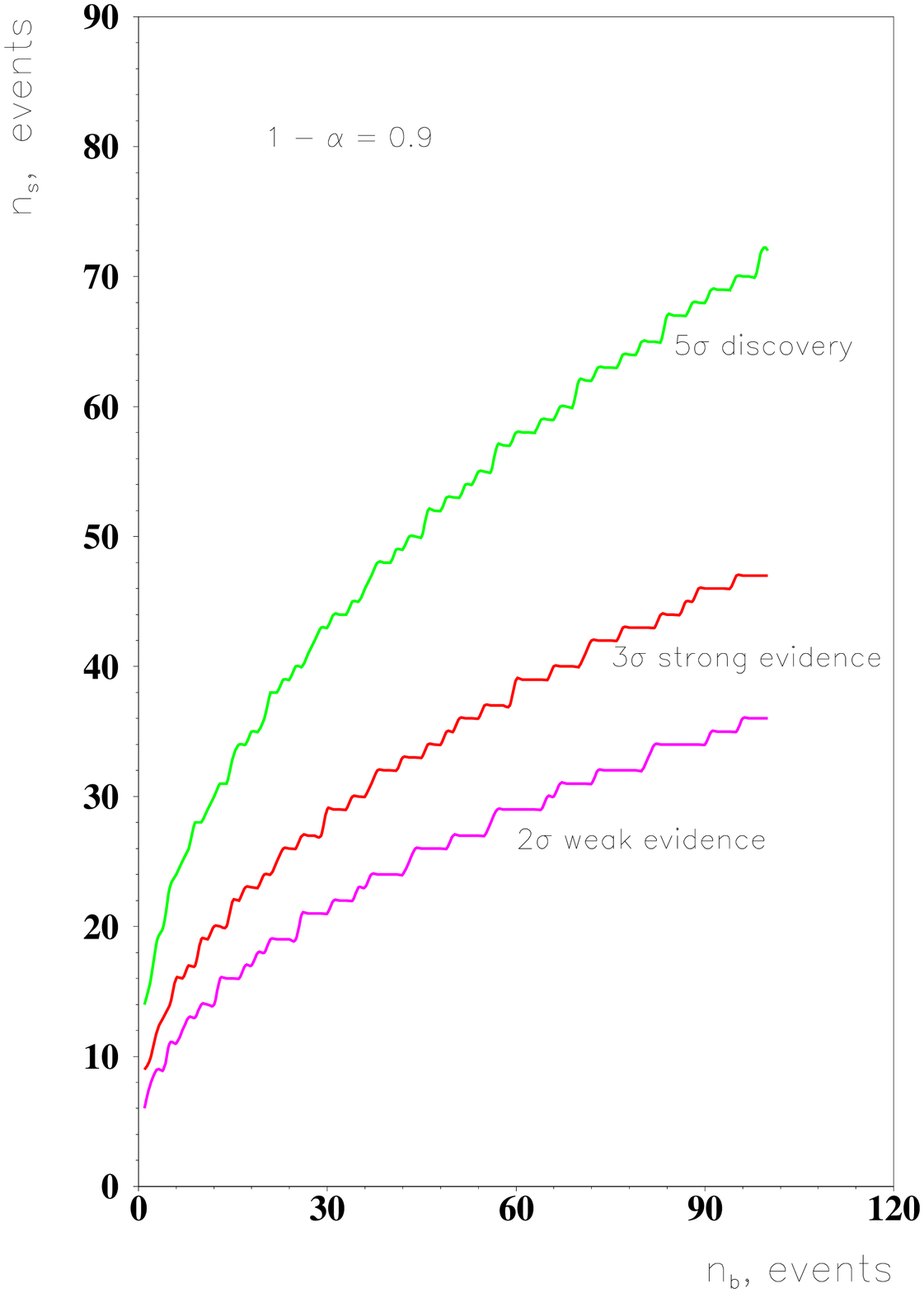,width=12.0cm}
\caption{Dependences $n_s$ versus $n_b$ for
$1 - \alpha = 0.9$ and for different values of $\beta$.}
    \label{fig:9} 
  \end{center}
\end{figure}
                
\begin{figure}[htpb]
  \begin{center}
  \epsfig{file=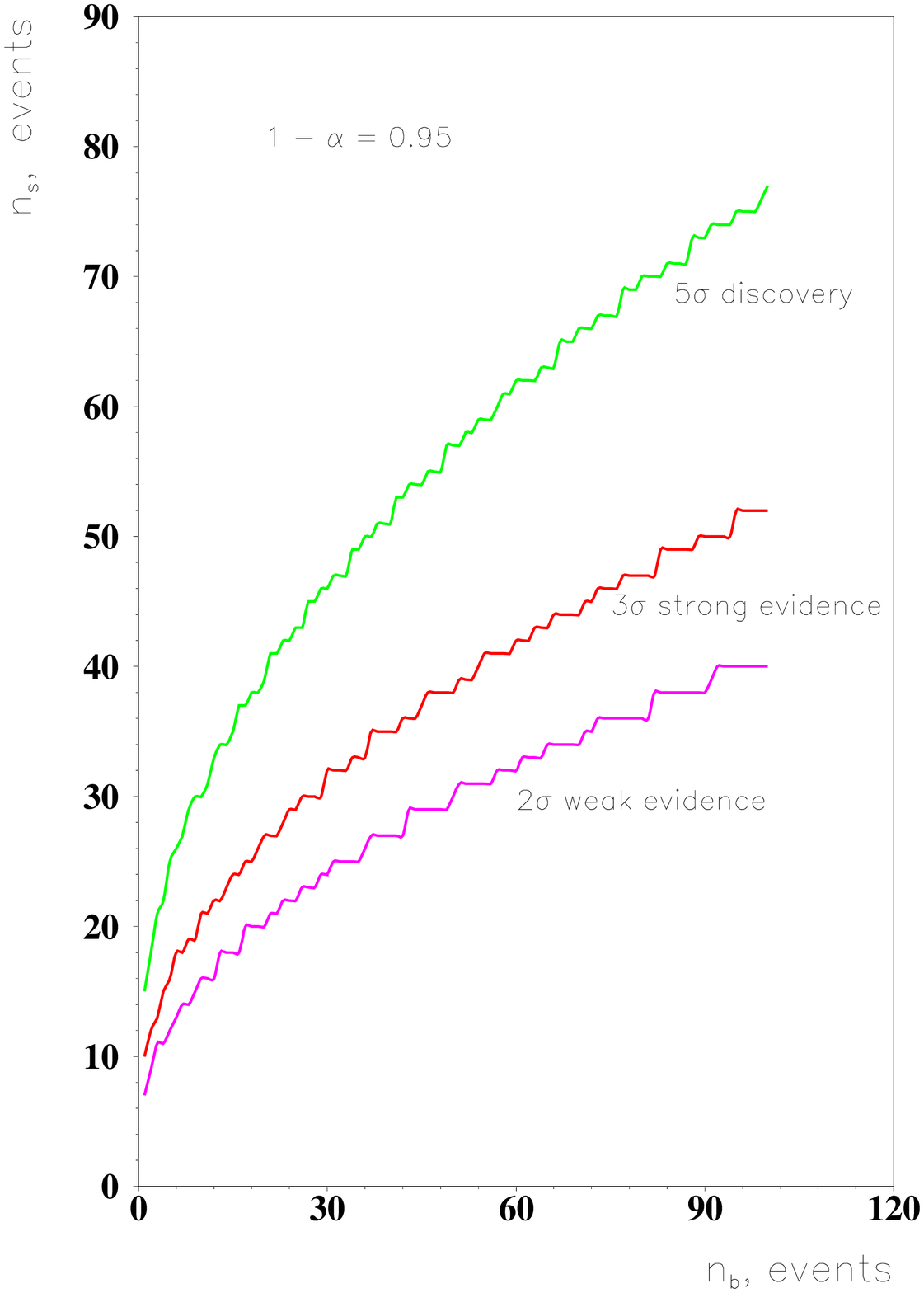,width=12.0cm}
\caption{Dependences $n_s$ versus $n_b$ for
$1 - \alpha = 0.95$ and for different values of $\beta$.}
    \label{fig:10} 
  \end{center}
\end{figure}
                
As an example consider the search for standard Higgs boson 
with a mass $m_h = 110~GeV$ using the $h \rightarrow \gamma \gamma $ 
decay mode at the CMS 
detector. For total luminosity $ L = 3 \cdot 10^{4} pb^{-1} ( 2 \cdot 
10^{4} pb^{-1})$ one can find \cite{CMS} that 
$n_b = 2893(1929), n_s = 357(238)$,
$\displaystyle S_1 = \frac{n_s}{\sqrt{n_b}} =6.6(5.4)$.
Using the formula (19) and Table of the standard normal probability 
density function~\cite{Frodesen} we find that 
$1 - \alpha(\Delta_{dis}) = 0.93(0.60)$. It means that for total 
luminosity $L = 3\cdot 10^{4}pb^{-1}(2\cdot 10^{4}pb^{-1})$ the CMS experiment 
will discover at $\geq 5\sigma$ level 
standard Higgs boson with a mass $m_h~=~110~GeV$ with a probability 93(60) 
percents \footnote{In other words let us suppose 
that we have constructed 100 identical 
CMS detectors. At $\geq 5\sigma$ level the Higgs boson will be discovered 
at 93(60) CMS detectors}. 

For the case when we are interested in estimation of the lower bound on
number $n_s$ of 
signal events (bound on new physics) we can use the equations     

\begin{equation}
1 - \alpha(\Delta) = \sum ^{\infty}_{n = n_0(\Delta) + 1}
f(n; n_b + n_s)
\end{equation}

\begin{equation}
\beta(\Delta) =  \sum ^{\infty}_{n=n_0({\Delta})+1} f(n; n_b) \geq \Delta
\end{equation}

\section{Exclusion limits.}

It is important to know the range in which the planned experiment can exclude 
the presence of signal at the given confidence level ($1 - \epsilon$).
It means that we will have uncertainty in future hypotheses testing
about non observation of signal equals to or less than $\epsilon$.
In refs.\cite{Hern,Taba} different methods to derive exclusion limits in 
prospective studies have been suggested. 
As is seen from Fig.11 the essential differences 
in values of the exclusion limits take place. Let us compare these 
methods by the use of the equal probability test~\cite{Bit2}.
In order to estimate 
the various approaches of the exclusion limit determination 
we suppose that new physics exists, i.e. the value $n_s$ equals 
to one of the exclusion limits from Fig.11 
and the value $n_b$ equals to the corresponding value of expected background.
Then we apply the equal probability test 
($f(n_0; n_s+n_b) = f(n_0; n_b)$) to find critical value $n_0$
for hypotheses testing in planned measurements (Fig.12). 
Here a zero hypothesis $H_0$ is 
the statement that new physics exists and an alternative hypothesis $H_1$ is 
the statement that new physics is absent.
After the calculation of the Type I error
$\alpha$ (the probability that the number of the observed events will be equal 
or less than the critical value $n_0$)
and the Type II error $\beta$ (the probability that the number of 
the observed events will bigger 
than the critical value $n_0$
in the case of the absence of new physics) we can compare the methods. 
For this purpose the relative uncertainty~\cite{Bit2}~~
$\tilde \kappa = \displaystyle \frac{\alpha+\beta}{2 - (\alpha+\beta)}$ which 
will take place under hypotheses testing $H_0$ versus $H_1$ is calculated.
This relative uncertainty $\tilde \kappa$ in case of applying the 
equal-probability test is a minimal relative value of the number of wrong 
decisions in the future hypotheses testing for Poisson distributions. It 
is the uncertainty in the observability of the new phenomenon. 
Note the $1 - \tilde \kappa$ (the relative number of correct decisions)
is a distance between two distributions
(the measure of distinguishability of two Poisson processes) 
in frequentist sense. 

\begin{figure}[htpb]
  \begin{center}
  \epsfig{file=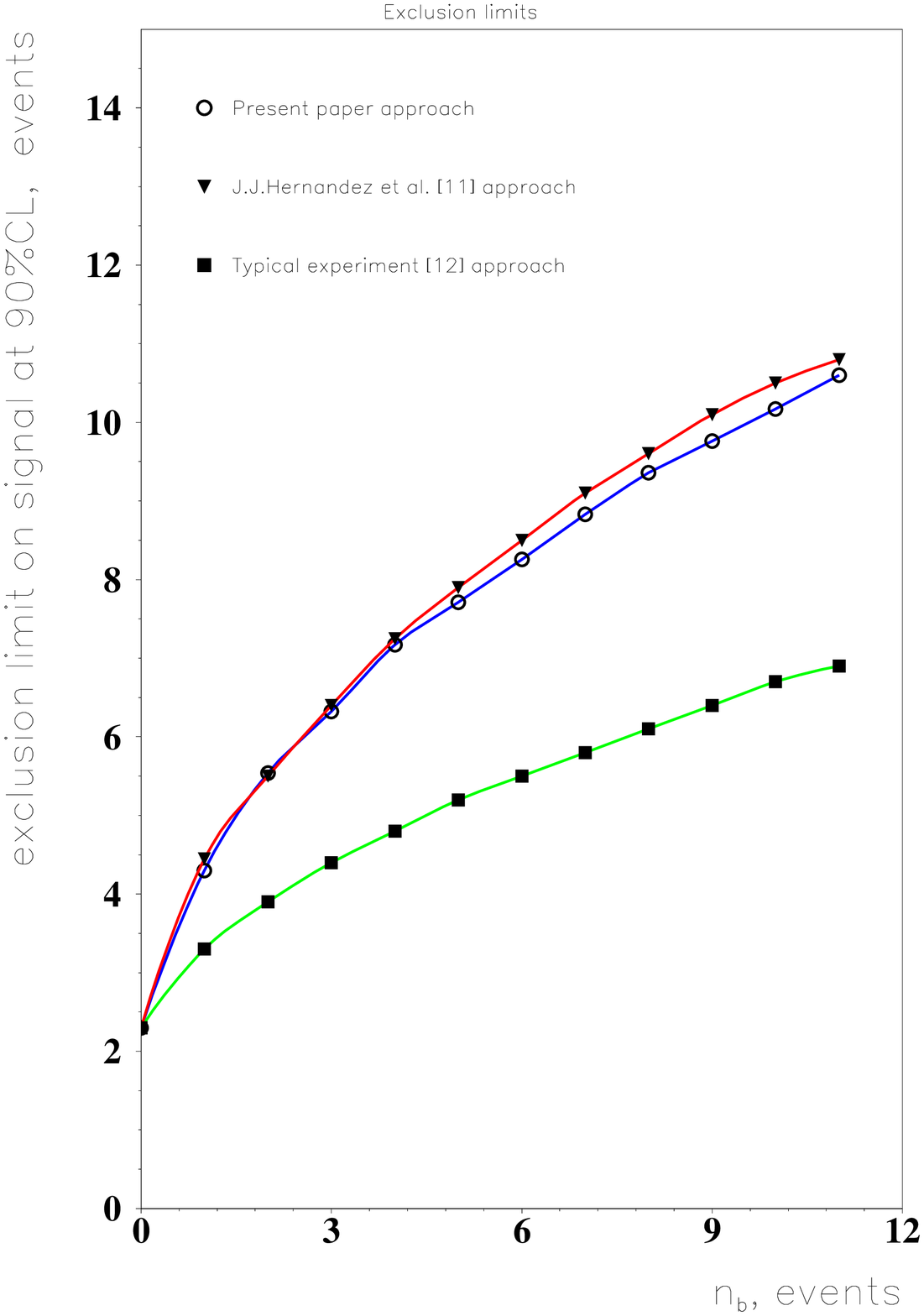,width=6.0cm}
\caption{Estimations of the 90\%~CL exclusion limit on the signal in a future 
experiment as a function of the expected background. The method proposed
in ref.~\cite{Taba} gives the values of exclusion limit 
close to "Typical experiment" approach.}
    \label{fig:11} 
  \end{center}

  \begin{center}
  \epsfig{file=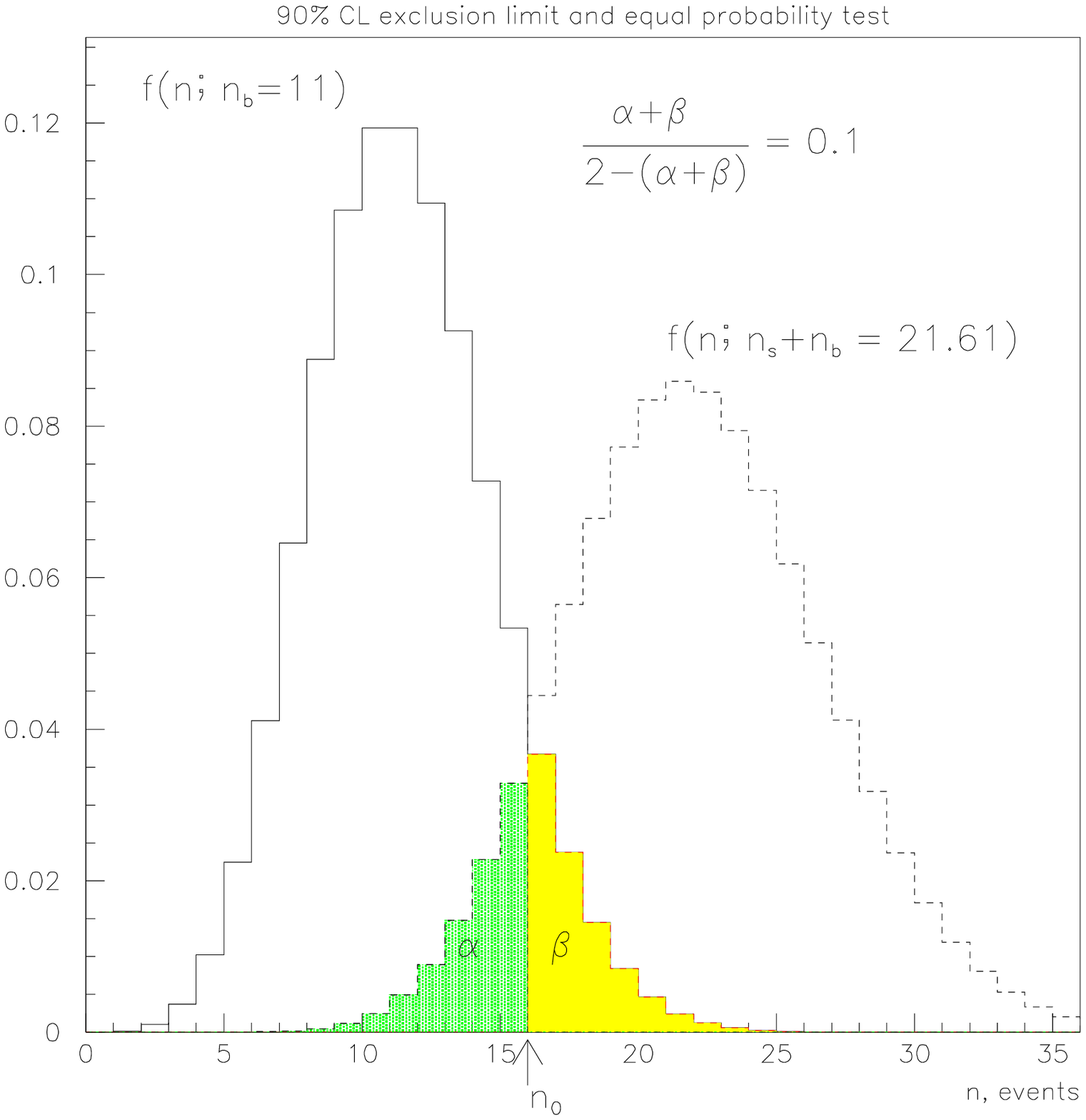,width=5.5cm}
\caption{Equal probability test for the case $n_b=11$ and $n_s=10.61$ 
gives the critical value $n_0=16$ and, correspondingly,   
the relative uncertainty $\tilde \kappa = 0.1$. It means if we observed
$n_{obs} \le n_0$~events in planned experiment we exclude the signal 
with average $n_s=10.61$ and above at 90\%CL.}
    \label{fig:12} 
  \end{center}
\end{figure}

In Table 1 the result of the comparison is shown. As is seen
from this Table the "Typical experiment" approach~\cite{Taba} gives
too small values of the exclusion limit. 
The difference in the 90\%~CL definition
is the main reason of the difference between our results and the 
results from ref.~\cite{Hern}. We require that 
$\tilde \kappa$ equals to $\epsilon$, i.e. we use only one parameter 
($\epsilon$) as measure of uncertainty in the hypotheses testing.
In ref~\cite{Hern} the criterion 
$\beta < \Delta$ and $\displaystyle \frac{\alpha}{1-\beta} < \epsilon$ 
for determination of the exclusion limits has been applied. 
It means the experiment will observe with probability at least $1 - \Delta$ 
at most a number of events such that the limit obtained at the 
$1 - \epsilon$ confidence level excludes the corresponding signal. 
In this case two parameters ($\epsilon,~\Delta$) are required to construct 
exclusion limits. Nevertheless, we have got results close 
to~\cite{Hern}~\footnote{The using $\kappa = \alpha + \beta$ as measure
of uncertainty~\cite{Bit2} gives a somewhat different results.}.
                                
\section{An account of systematic uncertainties related to nonexact 
knowledge of background and signal cross sections}

In  the previous sections we took into account only statistical 
fluctuation in the number of signal and background events and did 
not take into account other uncertainties. 
There is considered  
the systematic uncertainties~\footnote{In ref.~\cite{Cousins1} 
the systematic uncertainty is the uncertainty in the sensitivity factor.
This uncertainty has statistical properties which can be measured or
estimated. The systematic effects 
in ref.~\cite{DAgostini} as supposed has stochastic behaviour too. 
The account for statistical uncertainties due to statistical errors 
in determination of values $n_b$ and $n_s$~\cite{Bit4} implies 
the existence of conditional probability for parameter of Poisson 
distribution.
We consider here forthcoming experiments to search for new physics. 
In this case
the systematic uncertainties has theoretical origin without any statistical 
properties. }
due to imperfect knowledge of the 
background and signal cross sections.
 
In this section we investigate the influence of the systematic uncertainties 
related to nonexact knowledge of the background and signal cross sections 
on the discovery potential in planned experiments. 
Denote the Born background and signal cross sections as $\sigma^0_b$ and 
$\sigma^0_s$. An account of one loop corrections leads to 
$\sigma^0_b \rightarrow \sigma^0_b( 1 + \delta_{1b})$  and   
$\sigma^0_s \rightarrow \sigma^0_s( 1 + \delta_{1s})$, where 
typically $\delta_{1b}$ and $\delta_{1s}$ are $O(0.5)$ for the LHC.  
Two loop corrections 
for most reactions at present are not known. So, we can assume that the 
uncertainty related with nonexact knowledge of cross sections is 
around $\delta_{1s}$ and $\delta_{1b}$ correspondingly. In other words 
we assume that exact cross sections lie in the intervals 
$(\sigma_b, \sigma_b(1 +\delta_b))$ and  
$(\sigma_s, \sigma_s(1 +\delta_s))$, where $\sigma_b$ and $\sigma_s$
are calculated at Born or one loop level of the accuracy. 
The average number of background and 
signal events lie in the intervals $(n_b, n_b(1 + \delta_b))$, 
$(n_s, n_s(1 + \delta_s))$,
where $n_b = \sigma_b \cdot L$, $n_s = \sigma_s \cdot L$. 

To determine the new physics discovery potential we again have to 
compare two Poisson distributions with and without new physics. 
Contrary to the Section 3 we have to compare the Poisson 
distributions in which the average numbers lie in some intervals.  
It means that we have to find the critical value $n_0$ and to estimate 
the influence of systematic uncertainty on the discovery probability.  
A priori the only thing we know is that the average number of background 
and signal events lie in some intervals but we do not know the exact values 
of the average background and signal events. Moreover we can not say
anything about probability distributions of possible values $n_b$ and $n_s$
in this interval. Such distribution is absent.  

An account of uncertainties related to nonexact knowledge of background
and signal cross sections is straightforward and it is based on the 
results of Section 3. Suppose uncertainty in the calculation of exact 
background cross section is determined by parameter $\delta$, i.e. the exact 
cross section lies in the interval $(\sigma_b, \sigma_b (1+\delta))$ 
and the exact value of average number of background events 
lies in the interval 
$(n_b, n_b (1+\delta))$. Let us suppose $n_b \gg n_s$. In this instance
the discovery potential most sensitive to the systematic uncertainties.
Because we know nothing about possible values of average number of 
background events, we consider the worst case. Taking into account 
formulae (12) and (13) we have the formulae \footnote{Formulae (23,24) 
realize the worst case when the background 
cross section $ \sigma_b(1 +\delta)$ 
 is the maximal one, but we think that both the signal and the background 
cross sections are minimal}

\begin{equation}
\beta(\Delta) =  \sum ^{\infty}_{n=n_0({\Delta})+1} f(n; n_b(1+\delta)) 
\leq \Delta
\end{equation} 

\begin{equation}
1 - \alpha(\Delta) = \sum ^{\infty}_{n = n_0(\Delta) + 1}
f(n; n_b + n_s)
\end{equation}

This approach allows estimate the scale of influence of background
uncertainty to observability of signal.
As an application of formulae (23,24) consider the case
$n_b = n_s = 100$ (typical case for the search for 
supersymmetry at LHC). For such values of $n_s$ and $n_b$
and for  $\delta$= 0.,
0.1, 0.25, 0.5 we find that $1-\alpha(\Delta_{dis})$ = 0.9996,
0.9924, 0.8476, 0.137, correspondingly. So, we see that the uncertainty
in the calculations of background cross section is extremely essential for the
determination of the LHC discovery potential. Some other examples
are presented in Tables 2-7 and in Figs.13-14.

\begin{figure}[htpb]
  \begin{center}
  \epsfig{file=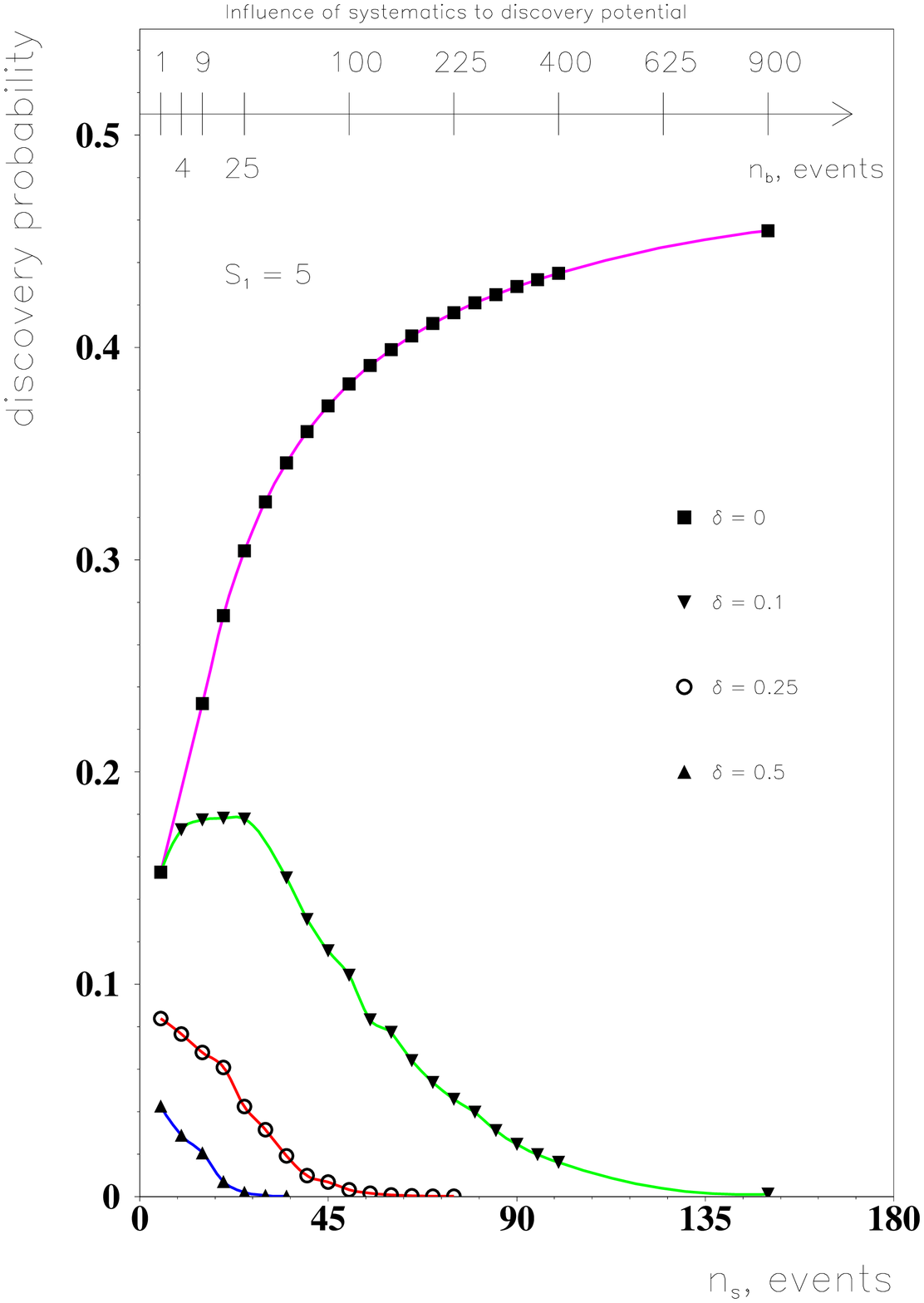,width=5.5cm}
\caption{Discovery probability versus $n_s$ for different values of
systematic uncertainty $\delta$ for the case $S_1=5$.}
    \label{fig:13} 
  \end{center}
                                
  \begin{center}
  \epsfig{file=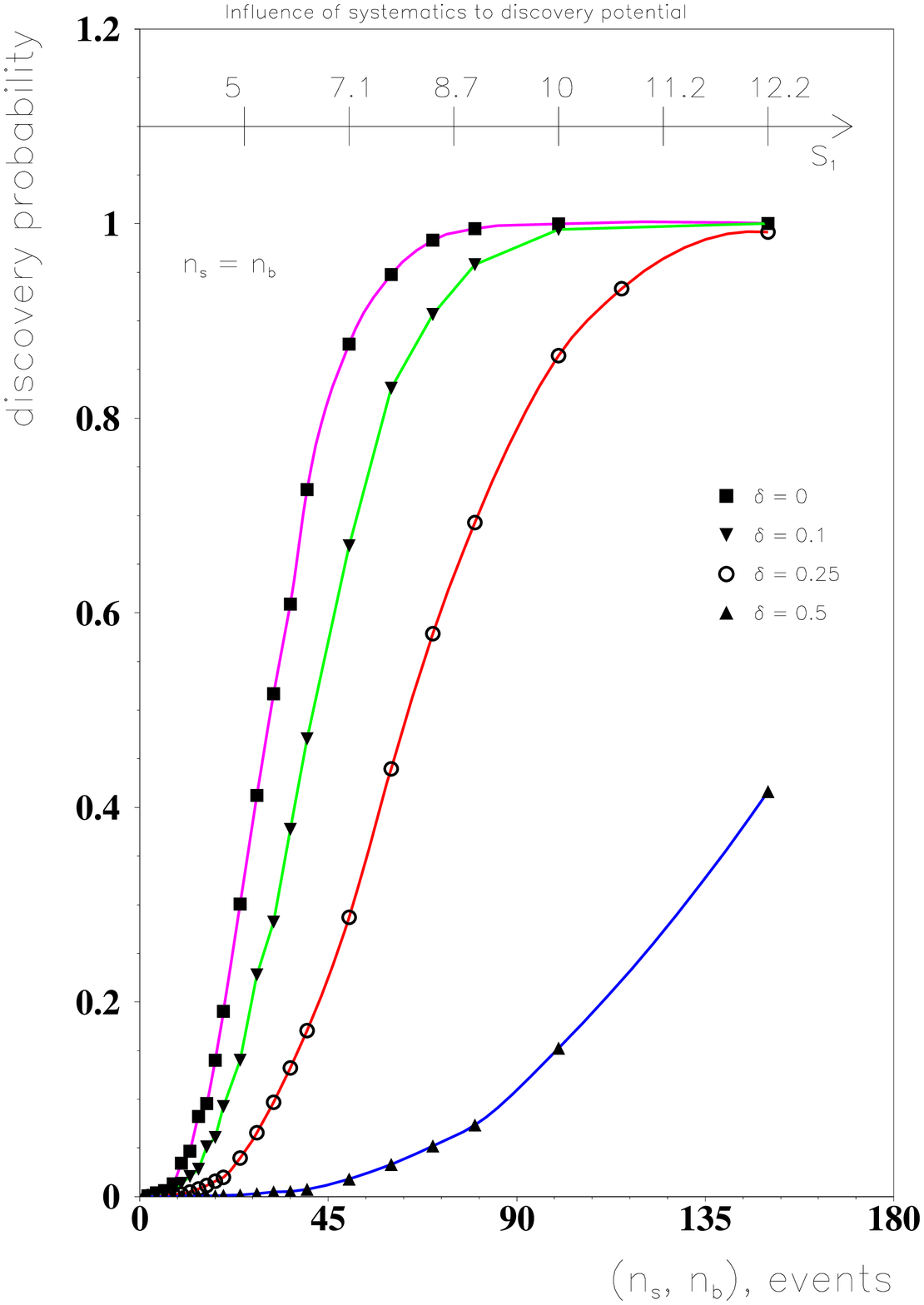,width=5.5cm}
\caption{Discovery probability versus $n_s$ for different values of
systematic uncertainty $\delta$ for the case $n_s=n_b$.}
    \label{fig:14} 
  \end{center}
\end{figure}

\section{Conclusions}

In this paper we have described  a method  to estimate  the discovery 
potential  and exclusion limits on new physics 
in planned experiments where only the average number of background $n_b$ and 
signal events  $n_s$ is known. We have found that in this case
the more proper definition of the significance (for $\alpha=0.5$) is 
$2 \cdot S_{12} = 2(\sqrt{n_s+n_b} - \sqrt{n_b)}$  
in comparison with often used expressions for the significances 
$S_1 = \displaystyle \frac{n_s}{\sqrt{n_b}}$ and
$S_2 = \displaystyle \frac{n_s}{\sqrt{n_s + n_b}}$.
For $1 - \alpha > 0.5$ we have additional additive contribution to the 
significance (see approximate formulae 19-20 of estimation of 
discovery potential for given discovery probability). As a result,
the effective significance $s$ of signal for given probability of
observation is proposed.
The results of direct calculations of dependences $n_s$ versus $n_b$
for different discovery probabilities and significances are presented.
We estimate the influence of systematic uncertainty related to nonexact 
knowledge of signal and background cross sections on the probability to 
discover new physics in planned experiments.
An account of such kind of systematics is very essential in the search 
for supersymmetry and leads to an essential decrease in the probability to 
discover new physics in future experiments. 
The texts of programs can be found in 
{\bf http://home.cern.ch/bityukov}.

We are grateful to Fred James, V.A.Matveev and V.F.Obraztsov for 
the interest and useful comments. 
S.B. thanks Bob Cousins, George Kahrimanis, James Linnemann, Louis Lyons,
Tony Vaiciulis, Alex Read, Byron Roy and Pekka Sinervo 
for very useful discussions. S.B. would like to thank James Stirling,
Mike Whalley and Linda Wilkinson for having organized this interesting
Conference which is the wonderful opportunity for exchange of ideas.
This work has been supported by CERN-INTAS 99-0377. 
The work of S.B. is also supported by CERN-INTAS 00-0440.

\newpage

\begin{table}
    \begin{center}
    \caption{The comparison of the different approaches to determination
of the exclusion limits. The $\alpha$ and the $\beta$ are the Type I and 
the Type II errors under the equal probability test. 
The $\tilde \kappa$ equals to 
$\displaystyle \frac{\alpha + \beta}{2 - (\alpha + \beta)}$.}
    \label{tab:Tab.1}
\begin{tabular}{|r|rrrr|rrrr|rrrr|}
\hline
     &  &this&paper & & & ref.& [11] & & &ref.&[12]&  \\ 
\hline
$n_b$&$n_s$&$\alpha$&$\beta$&$\tilde \kappa$&
$n_s$&$\alpha$&$\beta$&$\tilde \kappa$
&$n_s$&$\alpha$&$\beta$&$\tilde \kappa$\\ 
\hline
  1& 4.31&0.10 &0.08 &0.10 & 4.45 &0.09 &0.08 &0.09  &3.30 &0.20 &0.08 &0.16\\
  2& 5.54&0.13 &0.05 &0.10 & 5.50 &0.13 &0.05 &0.10  &3.90 &0.16 &0.14 &0.18\\
  3& 6.32&0.10 &0.08 &0.10 & 6.40 &0.09 &0.08 &0.10  &4.40 &0.14 &0.18 &0.19\\
  4& 7.19&0.13 &0.05 &0.10 & 7.25 &0.13 &0.05 &0.10  &4.80 &0.23 &0.11 &0.20\\
  5& 7.71&0.11 &0.07 &0.10 & 7.90 &0.10 &0.07 &0.09  &5.20 &0.20 &0.13 &0.20\\
  6& 8.26&0.10 &0.08 &0.10 & 8.41 &0.09 &0.08 &0.10  &5.50 &0.19 &0.15 &0.20\\
  7& 8.83&0.08 &0.10 &0.10 & 9.00 &0.08 &0.10 &0.10  &5.90 &0.17 &0.17 &0.20\\
  8& 9.36&0.12 &0.06 &0.10 & 9.70 &0.10 &0.06 &0.09  &6.10 &0.17 &0.18 &0.21\\
  9& 9.76&0.11 &0.07 &0.10 &10.16 &0.09 &0.07 &0.09  &6.40 &0.16 &0.20 &0.22\\
 10&10.17&0.10 &0.08 &0.10 &10.50 &0.09 &0.08 &0.09  &6.70 &0.22 &0.14 &0.22\\
 11&10.61&0.08 &0.11 &0.10 &10.80 &0.08 &0.09 &0.10  &6.90 &0.21 &0.15 &0.22\\
\hline
\end{tabular}
    \end{center}
\end{table}

\begin{table}[h]
    \caption{The dependence of $1-\alpha(\Delta_{dis})$ on
$n_s$ and $n_b$ for $S_1=5$ and different values of $\delta$.
$\Delta_{dis} = 2.85 \cdot 10^{-7}$.}
    \label{tab:Tab.2}
    \begin{center}
\begin{tabular}{|r|r|l|l|l|l|}
\hline
$n_s$&$n_b$&$\delta=0.0$&$\delta=0.1$&$\delta=0.25$&$\delta=0.5$\\ 
\hline
   5  &    1 &   0.1528  &     0.1528  &     0.0839   &    0.0426\\
  10  &    4 &   0.2441  &     0.1728  &     0.0765   &    0.0288\\
  15  &    9 &   0.2323  &     0.1775  &     0.0678   &    0.0206\\
  20  &   16 &   0.2737  &     0.1783  &     0.0609   &    0.0071\\
  25  &   25 &   0.3041  &     0.1779  &     0.0424   &    0.0020\\
  30  &   36 &   0.3273  &     0.1480  &     0.0315   &    0.0005\\
  35  &   49 &   0.3456  &     0.1502  &     0.0192   &    0.0001\\
  40  &   64 &   0.3603  &     0.1305  &     0.0098   &          \\
  45  &   81 &   0.3725  &     0.1157  &     0.0068   &           \\
  50  &  100 &   0.3828  &     0.1042  &     0.0032   &  \\
  55  &  121 &   0.3915  &     0.0833  &     0.0015   &  \\
  60  &  144 &   0.3990  &     0.0773  &     0.0008   &  \\
  65  &  169 &   0.4055  &     0.0640  &     0.0004   &  \\
  70  &  196 &   0.4113  &     0.0538  &     0.0002   &  \\
  75  &  225 &   0.4163  &     0.0459  &     0.0001   &  \\
  80  &  256 &   0.4209  &     0.0397  &              &  \\
  85  &  289 &   0.4249  &     0.0310  &              &  \\
  90  &  324 &   0.4286  &     0.0246  &              &  \\
  95  &  361 &   0.4319  &     0.0197  &              &  \\
 100  &  400 &   0.4350  &     0.0161  &              &  \\
 150  &  900 &   0.4550  &     0.0011  &              &  \\
\hline
\end{tabular}
    \end{center}
\end{table}

\begin{table}[h]
    \caption{The dependence of $1-\alpha(\Delta_{dis})$ on
$n_s$ and $n_b$ for $S_2 \approx 5$ and different values of $\delta$.}
    \label{tab:Tab.3}
    \begin{center}
\begin{tabular}{|r|r|l|l|l|l|}
\hline
$n_s$&$n_b$&$\delta=0.$&$\delta=0.1$&$\delta=0.25$&$\delta=0.5$\\ 
\hline
  26 &     1 &   1.0000 &      1.0000  &     0.9999  &     0.9998    \\
  29 &     4 &   0.9992 &      0.9983  &     0.9940  &     0.9825    \\
  33 &     9 &   0.9909 &      0.9856  &     0.9524  &     0.8786    \\
  37 &    16 &   0.9725 &      0.9473  &     0.8491  &     0.5730    \\
  41 &    25 &   0.9418 &      0.8806  &     0.6606  &     0.2457    \\
  45 &    36 &   0.9016 &      0.7622  &     0.4705  &     0.0696 \\
  50 &    49 &   0.8774 &      0.7058  &     0.3208  &     0.0222 \\
  55 &    64 &   0.8546 &      0.6206  &     0.1909  &     0.0044 \\
 100 &   300 &   0.6803 &      0.1110  &     0.0001  & \\
 150 &   750 &   0.6224 &      0.0084  &             & \\
\hline
\end{tabular}
    \end{center}
\end{table}

\begin{table}[h]
  \caption{$n_s = \frac{1}{5} \cdot n_b$. The dependence of 
$1-\alpha(\Delta_{dis})$ 
on $n_s$ and $n_b$ for different values of~$\delta$.}
    \label{tab:Tab.4}
    \begin{center}
\begin{tabular}{|r|r|l|l|}
\hline
$n_s$&$n_b$&$\delta=0.$&$\delta=0.1$\\ 
\hline
     50 &   250  &  0.0319  &     0.0003  \\ 
    100 &   500  &  0.2621  &     0.0023  \\ 
    150 &   750  &  0.6224  &     0.0084  \\ 
    200 &  1000  &  0.8671  &     0.0232  \\ 
    250 &  1250  &  0.9644  &     0.0513  \\ 
    300 &  1500  &  0.9926  &     0.0920  \\     
    350 &  1750  &  0.9988  &     0.1500  \\    
    400 &  2000  &  0.9998  &     0.2156  \\     
\hline
\end{tabular}
    \end{center}
\end{table}

\begin{table}[h]
  \caption{$n_s = \frac{1}{10} \cdot n_b$. The dependence 
of $1-\alpha(\Delta_{dis})$ on $n_s$ and $n_b$.}
    \label{tab:Tab.5}
    \begin{center}
\begin{tabular}{|r|r|l|}
\hline
$n_s$&$n_b$&$\delta=0.$\\ 
\hline
     50 &   500 &   0.0030  \\
    100 &  1000 &   0.0327  \\
    150 &  1500 &   0.1214  \\    
    200 &  2000 &   0.2781  \\    
    250 &  2500 &   0.4721  \\    
    300 &  3000 &   0.6514  \\    
    350 &  3500 &   0.7919  \\    
    400 &  4000 &   0.8878  \\    
\hline
\end{tabular}
    \end{center}
\end{table}

\begin{table}[h]
    \caption{$n_s$ = $n_b$. The dependence of $1-\alpha(\Delta_{dis})$ on
$n_s$ and $n_b$ for different values of $\delta$.}
    \label{tab:Tab.6}
    \begin{center}
\begin{tabular}{|r|r|l|l|l|l|}
\hline
$n_s$&$n_b$&$\delta=0.$&$\delta=0.1$&$\delta=0.25$&$\delta=0.5$\\ 
\hline
      2  &    2 &   0.0009    &   0.0003    &   0.0001    &   0.00002\\
      4  &    4 &   0.0037    &   0.0016    &   0.0003    &   0.00003\\
      6  &    6 &   0.0061    &   0.0030    &   0.0007    &   0.00006\\
      8  &    8 &   0.0131    &   0.0075    &   0.0022    &   0.00013\\
     10  &   10 &   0.0343    &   0.0135    &   0.0027    &   0.0002\\
     12  &   12 &   0.0467    &   0.0206    &   0.0050    &   0.0003\\
     14  &   14 &   0.0822    &   0.0283    &   0.0080    &   0.0004\\
     16  &   16 &   0.0956    &   0.0512    &   0.0116    &   0.0007\\
     18  &   18 &   0.1401    &   0.0609    &   0.0156    &   0.0007\\
     20  &   20 &   0.1904    &   0.0925    &   0.0200    &   0.0012\\
     24  &   24 &   0.3005    &   0.1402    &   0.0395    &   0.0017\\
     28  &   28 &   0.4122    &   0.2280    &   0.0655    &   0.0031\\
     32  &   32 &   0.5166    &   0.2821    &   0.0969    &   0.0050\\
     36  &   36 &   0.6089    &   0.3773    &   0.1323    &   0.0054\\
     40  &   40 &   0.7268    &   0.4703    &   0.1704    &   0.0076\\
     50  &   50 &   0.8762    &   0.6688    &   0.2872    &   0.0181\\
     60  &   60 &   0.9477    &   0.8309    &   0.4397    &   0.0332\\
     70  &   70 &   0.9831    &   0.9067    &   0.5784    &   0.0520\\
     80  &   80 &   0.9949    &   0.9575    &   0.6929    &   0.0737\\
    100  &  100 &   0.9997    &   0.9938    &   0.8641    &   0.1527\\  
    150  &  150 &   1.0000    &   1.0000    &   0.9914    &   0.4163\\  
\hline
\end{tabular}
    \end{center}
\end{table}

\begin{table}[h]
  \caption{$n_s~=~0.5 \cdot n_b$. The dependence of 
$1-\alpha(\Delta_{dis})$ on $n_s$ and $n_b$ for different 
values of $\delta$.}
    \label{tab:Tab.7}
    \begin{center}
\begin{tabular}{|r|r|l|l|l|}
\hline
$n_s$&$n_b$&$\delta=0.$&$\delta=0.1$&$\delta=0.25$\\ 
\hline
      2 &     4 &   0.0002 &      0.0001 &      0.0000 \\
      4 &     8 &   0.0003 &      0.0001 &      0.0000 \\
      6 &    12 &   0.0010 &      0.0002 &      0.0000 \\
      8 &    16 &   0.0017 &      0.0005 &      0.0000 \\
     10 &    20 &   0.0040 &      0.0009 &      0.0001 \\
     12 &    24 &   0.0071 &      0.0012 &      0.0001 \\
     14 &    28 &   0.0111 &      0.0023 &      0.0001 \\
     16 &    32 &   0.0156 &      0.0025 &      0.0002 \\
     18 &    36 &   0.0206 &      0.0039 &      0.0002 \\
     20 &    40 &   0.0341 &      0.0056 &      0.0003 \\
     24 &    48 &   0.0589 &      0.0099 &      0.0003 \\
     28 &    56 &   0.0885 &      0.0149 &      0.0005 \\
     32 &    64 &   0.1107 &      0.0259 &      0.0008 \\
     36 &    72 &   0.1796 &      0.0329 &      0.0013 \\
     40 &    80 &   0.2171 &      0.0482 &      0.0016 \\
     50 &   100 &   0.3828 &      0.1042 &      0.0032 \\
     60 &   120 &   0.5396 &      0.1753 &      0.0061 \\
     70 &   140 &   0.6947 &      0.2539 &      0.0099 \\
     80 &   160 &   0.8076 &      0.3578 &      0.0144 \\
    100 &   200 &   0.9311 &      0.5537 &      0.0319 \\
    150 &   300 &   0.9979 &      0.8861 &      0.1153 \\
\hline
\end{tabular}
    \end{center}
\end{table}

\end{document}